\RequirePackage{lineno}
\documentclass[aps,prl,twocolumn,showpacs,superscriptaddress,groupedaddress]{revtex4}  

\usepackage{graphicx}  
\usepackage{dcolumn}   
\usepackage{bm}        
\usepackage{amssymb}   

\usepackage{epsfig}
\usepackage{subfigure}
\usepackage{rotating}

\newcommand{\ttbar}{$t\overline{t}$}

\newcommand{\MET}{\mbox{$\not\!\!E_T$}}

\newcommand{\explimA} {4.7} 
\newcommand{\obslimA} {5.2} 

\newcommand{\diboexpsig}{1.8}
\newcommand{\dibosig}   {1.0}
\newcommand{\diboxsec}	{0.55 $\pm$ 0.36\,(stat.) $\pm$ 0.37\,(syst.)}
\newcommand{\diboSMxsec} {4.4~$\pm$~0.3~pb}

\begin{document}

\hspace{5.2in} \mbox{Fermilab-Pub-12-405-E}

\title{Search for the standard model Higgs boson in associated $\bm{WH}$ production in 9.7~fb$\bm{^{-1}}$ of $\bm{p\bar{p}}$ collisions with the D0 detector}
\affiliation{LAFEX, Centro Brasileiro de Pesquisas F\'{i}sicas, Rio de Janeiro, Brazil}
\affiliation{Universidade do Estado do Rio de Janeiro, Rio de Janeiro, Brazil}
\affiliation{Universidade Federal do ABC, Santo Andr\'e, Brazil}
\affiliation{University of Science and Technology of China, Hefei, People's Republic of China}
\affiliation{Universidad de los Andes, Bogot\'a, Colombia}
\affiliation{Charles University, Faculty of Mathematics and Physics, Center for Particle Physics, Prague, Czech Republic}
\affiliation{Czech Technical University in Prague, Prague, Czech Republic}
\affiliation{Center for Particle Physics, Institute of Physics, Academy of Sciences of the Czech Republic, Prague, Czech Republic}
\affiliation{Universidad San Francisco de Quito, Quito, Ecuador}
\affiliation{LPC, Universit\'e Blaise Pascal, CNRS/IN2P3, Clermont, France}
\affiliation{LPSC, Universit\'e Joseph Fourier Grenoble 1, CNRS/IN2P3, Institut National Polytechnique de Grenoble, Grenoble, France}
\affiliation{CPPM, Aix-Marseille Universit\'e, CNRS/IN2P3, Marseille, France}
\affiliation{LAL, Universit\'e Paris-Sud, CNRS/IN2P3, Orsay, France}
\affiliation{LPNHE, Universit\'es Paris VI and VII, CNRS/IN2P3, Paris, France}
\affiliation{CEA, Irfu, SPP, Saclay, France}
\affiliation{IPHC, Universit\'e de Strasbourg, CNRS/IN2P3, Strasbourg, France}
\affiliation{IPNL, Universit\'e Lyon 1, CNRS/IN2P3, Villeurbanne, France and Universit\'e de Lyon, Lyon, France}
\affiliation{III. Physikalisches Institut A, RWTH Aachen University, Aachen, Germany}
\affiliation{Physikalisches Institut, Universit\"at Freiburg, Freiburg, Germany}
\affiliation{II. Physikalisches Institut, Georg-August-Universit\"at G\"ottingen, G\"ottingen, Germany}
\affiliation{Institut f\"ur Physik, Universit\"at Mainz, Mainz, Germany}
\affiliation{Ludwig-Maximilians-Universit\"at M\"unchen, M\"unchen, Germany}
\affiliation{Fachbereich Physik, Bergische Universit\"at Wuppertal, Wuppertal, Germany}
\affiliation{Panjab University, Chandigarh, India}
\affiliation{Delhi University, Delhi, India}
\affiliation{Tata Institute of Fundamental Research, Mumbai, India}
\affiliation{University College Dublin, Dublin, Ireland}
\affiliation{Korea Detector Laboratory, Korea University, Seoul, Korea}
\affiliation{CINVESTAV, Mexico City, Mexico}
\affiliation{Nikhef, Science Park, Amsterdam, Netherlands}
\affiliation{Radboud University Nijmegen, Nijmegen, Netherlands}
\affiliation{Joint Institute for Nuclear Research, Dubna, Russia}
\affiliation{Institute for Theoretical and Experimental Physics, Moscow, Russia}
\affiliation{Moscow State University, Moscow, Russia}
\affiliation{Institute for High Energy Physics, Protvino, Russia}
\affiliation{Petersburg Nuclear Physics Institute, St. Petersburg, Russia}
\affiliation{Instituci\'{o} Catalana de Recerca i Estudis Avan\c{c}ats (ICREA) and Institut de F\'{i}sica d'Altes Energies (IFAE), Barcelona, Spain}
\affiliation{Uppsala University, Uppsala, Sweden}
\affiliation{Lancaster University, Lancaster LA1 4YB, United Kingdom}
\affiliation{Imperial College London, London SW7 2AZ, United Kingdom}
\affiliation{The University of Manchester, Manchester M13 9PL, United Kingdom}
\affiliation{University of Arizona, Tucson, Arizona 85721, USA}
\affiliation{University of California Riverside, Riverside, California 92521, USA}
\affiliation{Florida State University, Tallahassee, Florida 32306, USA}
\affiliation{Fermi National Accelerator Laboratory, Batavia, Illinois 60510, USA}
\affiliation{University of Illinois at Chicago, Chicago, Illinois 60607, USA}
\affiliation{Northern Illinois University, DeKalb, Illinois 60115, USA}
\affiliation{Northwestern University, Evanston, Illinois 60208, USA}
\affiliation{Indiana University, Bloomington, Indiana 47405, USA}
\affiliation{Purdue University Calumet, Hammond, Indiana 46323, USA}
\affiliation{University of Notre Dame, Notre Dame, Indiana 46556, USA}
\affiliation{Iowa State University, Ames, Iowa 50011, USA}
\affiliation{University of Kansas, Lawrence, Kansas 66045, USA}
\affiliation{Kansas State University, Manhattan, Kansas 66506, USA}
\affiliation{Louisiana Tech University, Ruston, Louisiana 71272, USA}
\affiliation{Boston University, Boston, Massachusetts 02215, USA}
\affiliation{Northeastern University, Boston, Massachusetts 02115, USA}
\affiliation{University of Michigan, Ann Arbor, Michigan 48109, USA}
\affiliation{Michigan State University, East Lansing, Michigan 48824, USA}
\affiliation{University of Mississippi, University, Mississippi 38677, USA}
\affiliation{University of Nebraska, Lincoln, Nebraska 68588, USA}
\affiliation{Rutgers University, Piscataway, New Jersey 08855, USA}
\affiliation{Princeton University, Princeton, New Jersey 08544, USA}
\affiliation{State University of New York, Buffalo, New York 14260, USA}
\affiliation{University of Rochester, Rochester, New York 14627, USA}
\affiliation{State University of New York, Stony Brook, New York 11794, USA}
\affiliation{Brookhaven National Laboratory, Upton, New York 11973, USA}
\affiliation{Langston University, Langston, Oklahoma 73050, USA}
\affiliation{University of Oklahoma, Norman, Oklahoma 73019, USA}
\affiliation{Oklahoma State University, Stillwater, Oklahoma 74078, USA}
\affiliation{Brown University, Providence, Rhode Island 02912, USA}
\affiliation{University of Texas, Arlington, Texas 76019, USA}
\affiliation{Southern Methodist University, Dallas, Texas 75275, USA}
\affiliation{Rice University, Houston, Texas 77005, USA}
\affiliation{University of Virginia, Charlottesville, Virginia 22904, USA}
\affiliation{University of Washington, Seattle, Washington 98195, USA}
\author{V.M.~Abazov} \affiliation{Joint Institute for Nuclear Research, Dubna, Russia}
\author{B.~Abbott} \affiliation{University of Oklahoma, Norman, Oklahoma 73019, USA}
\author{B.S.~Acharya} \affiliation{Tata Institute of Fundamental Research, Mumbai, India}
\author{M.~Adams} \affiliation{University of Illinois at Chicago, Chicago, Illinois 60607, USA}
\author{T.~Adams} \affiliation{Florida State University, Tallahassee, Florida 32306, USA}
\author{G.D.~Alexeev} \affiliation{Joint Institute for Nuclear Research, Dubna, Russia}
\author{G.~Alkhazov} \affiliation{Petersburg Nuclear Physics Institute, St. Petersburg, Russia}
\author{A.~Alton$^{a}$} \affiliation{University of Michigan, Ann Arbor, Michigan 48109, USA}
\author{G.~Alverson} \affiliation{Northeastern University, Boston, Massachusetts 02115, USA}
\author{A.~Askew} \affiliation{Florida State University, Tallahassee, Florida 32306, USA}
% ...
%
%
\collaboration{D0 Collaboration\footnote{with visitors from
$^{a}$Augustana College, Sioux Falls, SD, USA,
$^{b}$The University of Liverpool, Liverpool, UK,
$^{c}$UPIITA-IPN, Mexico City, Mexico,
$^{d}$DESY, Hamburg, Germany,
$^{e}$SLAC, Menlo Park, CA, USA,
$^{f}$University College London, London, UK,
$^{g}$Centro de Investigacion en Computacion - IPN, Mexico City, Mexico,
$^{h}$ECFM, Universidad Autonoma de Sinaloa, Culiac\'an, Mexico
and
$^{i}$Universidade Estadual Paulista, S\~ao Paulo, Brazil.
}} \noaffiliation
\vskip 0.25cm
      
\date{\today}

\begin{abstract}
We present a search for the standard model Higgs boson in final states with a charged lepton (electron or muon), 
missing transverse energy, and two or three jets, at least one of which is identified as a $b$-quark jet.
The search is primarily sensitive to $WH\to\ell\nu b\bar{b}$
production and uses data corresponding to 9.7~fb$^{-1}$ of integrated luminosity collected
with the D0 detector at the Fermilab Tevatron $p\bar{p}$ Collider at $\sqrt{s}=1.96$~TeV.
We observe agreement between the data and the expected background.
For a Higgs boson mass of 125~GeV, we set a 95\%\ C.L. upper limit on the production of a standard model Higgs boson of
\obslimA$\times\sigma_{\rm SM}$, where $\sigma_{\rm SM}$ is the standard model Higgs boson production cross section, while the expected limit is 
\explimA$\times\sigma_{\rm SM}$.
\end{abstract}

\maketitle

The Higgs boson is the only fundamental particle in the standard model (SM) predicted as a direct consequence of the Higgs mechanism describing spontaneous electroweak symmetry  breaking \cite{Higgs:1964pj,Englert:1964et,Guralnik:1964eu}.

The Higgs mechanism generates the masses of the weak gauge bosons and provides an explanation for the nonzero masses of 
fermions generated by their Yukawa couplings to the Higgs field. 
The mass of the Higgs boson ($M_{H}$) is a free parameter in the SM that must be constrained by experimental results. 
The direct searches at the CERN $e^+e^-$ Collider (LEP)~\cite{Barate:2003sz} 
exclude $M_H < 114.4$~GeV at the 95\% confidence level (C.L.) and precision measurements of other electroweak parameters constrain $M_{H}$
to be less than $152$~GeV~\cite{Aaltonen:2012bp,Abazov:2012bv,bib:LEPEWWG}.
The region $147 < M_H < 179$~GeV is excluded by the combined analysis of the CDF and D0 Collaborations~\cite{CDFandD0:2012aa}. 
The ATLAS and CMS Collaborations at the CERN Large Hadron Collider (LHC) have excluded much of the allowed mass range
and reported excesses at the 2--3 standard deviation (s.d.) level for $M_H\approx 125$~GeV~\cite{Atlas-jul2012,CMS-PAS-HIG-12-008}.
The experiments now exclude $111< M_H < 122$~GeV, $129< M_H < 559$~GeV (ATLAS)~\cite{atlas-obs}, and $110< M_H < 122$~GeV, $127< M_H < 600$~GeV (CMS)~\cite{cms-obs}. 
Both experiments have observed a resonance consistent with SM Higgs production at $M_H\approx 125$~GeV,
primarily in the $\gamma\gamma$ and $ZZ$ final states, above the 5 s.d. level~\cite{atlas-obs, cms-obs}.
Demonstrating that the observed resonance is due to
SM Higgs boson production requires also observing it in the $b\bar{b}$ final state, which is the dominant decay mode in this mass range.

The dominant Higgs boson production process at the Tevatron Collider is gluon-gluon fusion. 
The associated production of a Higgs boson with a weak boson occurs at a rate about 3 times lower than the gluon-gluon fusion production process but is of particular importance in Higgs boson searches.
At masses below $M_H \approx 135$~GeV,
$H \rightarrow b \bar{b}$ decays dominate
but are difficult to distinguish from background when the Higgs boson is produced by gluon-gluon fusion.
Instead, associated production of a Higgs boson and a $W$ boson
is one of the most sensitive search channels at the Tevatron. 

This Letter presents a search based on events with 
one charged lepton ($\ell=e$ or $\mu$), 
an imbalance in transverse energy (\MET) that arises from the
neutrino in the $W\to\ell\nu$ decay, and two or three jets, 
where one or more of these jets is selected as a candidate $b$ quark (``$b$-tagged'') jet.
The search is also sensitive to $ZH$ production when one of the charged leptons
from the $Z\to\ell^+\ell^-$ decay is not identified.
The analysis is optimized by subdividing into channels with
different background compositions and signal to background ratios
based on lepton flavor, jet multiplicity, and the number and quality of candidate $b$-quark jets.

Several searches for $WH \to \ell\nu b\bar{b}$ production have already been reported at a $p\bar{p}$ 
center-of-mass energy of $\sqrt{s}=1.96$~TeV, most recently 
by the CDF Collaboration \cite{Aaltonen:2012wh}.
Previous searches \cite{Abazov:2005aa, Abazov:2007hk, Abazov:2008eb, Abazov:2010hn, Abazov:2012wh} by the D0 Collaboration
use subsamples of the data presented in this Letter with integrated luminosities up to 5.3 fb$^{-1}$.
We present an updated search using a multivariate approach 
with a full dataset which, after imposing data quality requirements, corresponds to an integrated luminosity of 9.7 fb$^{-1}$.

This analysis uses most of the major components of the D0 detector, described in detail in Refs.~\cite{Abachi:1993em,Abazov:2005pn,Abolins:2007yz,Angstadt:2009ie}.
Events in the electron channel are selected with
triggers requiring an electromagnetic object in the calorimeter or an electromagnetic object with additional
jets. In the muon channel we use a mixture of single muon, muon
plus jet, \MET\ plus jet, and multijet triggers. We
correct simulated events for trigger efficiency by using a
method similar to that described in Ref.~\cite{Abazov:2012wh}.

Several SM processes produce or can mimic a final state with a charged lepton,
\MET, and jets, including 
diboson ($WW$, $WZ$, and $ZZ$), $V$+jets ($V=W$~or~$Z$), $t\bar{t}$, single top quark, and multijet (MJ) production. We estimate the MJ 
background from data and other backgrounds from simulation.
The $V$+jets and \ttbar\ samples are simulated with the {\sc alpgen}~\cite{Mangano:2002ea} Monte Carlo (MC) generator
interfaced to {\sc pythia}~\cite{Sjostrand:2006za} for parton showering and hadronization. {\sc alpgen} samples are produced by
using the MLM parton-jet matching prescription~\cite{Mangano:2002ea}. The $V$+jets samples contain $V+jj$ (where $j=u,d,s,$ or $g$) and $V+cj$ (together denoted as ``$V$+light-flavor'') processes, and
$V+b\bar{b}$ and $V+c\bar{c}$ (together denoted as ``$V$+heavy-flavor''), generated separately from $V$+light-flavor.
{\sc pythia} is used to simulate the production of dibosons ($WW$, $WZ$, and $ZZ$) and all signal processes.
Single top quark events
are generated with the {\sc singletop} event generator~\cite{Boos:2004kh,Boos:2006af} using {\sc pythia} for parton evolution and hadronization.
Simulation of background and signal processes uses the CTEQ6L1~\cite{Lai:1996mg,Pumplin:2002vw} 
leading-order (LO) parton distribution functions.
Events are processed through a full D0 detector simulation based on {\sc geant}~\cite{geant}.
To account for multiple $p\bar{p}$ interactions,
all generated events are overlaid with an event from a sample of random beam
crossings with the same
instantaneous luminosity profile as the data. Further on, events are reconstructed by using the same software as is used for the data.

The signal cross sections and branching fractions $\mathcal{B}$ are
normalized to the SM predictions~\cite{CDFandD0:2012aa}.
Next-to-LO (NLO) cross sections are used for 
single top quark~\cite{Kidonakis:2006bu}
and diboson~\cite{Campbell:1999ah,mcfm_code} production and
approximate next-to-NLO (NNLO) for \ttbar\ production~\cite{Langenfeld:2009wd}.
The $V$+jets processes are normalized to the NNLO cross section~\cite{Hamberg:1990np} 
with MSTW2008 NNLO parton distribution functions~\cite{Martin:2004ir}.
The $V$+heavy-flavor events are corrected by using the 
NLO to LO ratio obtained from the Monte Carlo program {\sc mcfm}~\cite{Campbell:2001ik,mcfm_code}.
We compare the data with the prediction for $V$+jets production
and find a relative data to MC normalization factor 
of $1.0 \pm 0.1$, obtained after subtracting all other expected
background processes and before $b$ tagging.

This analysis begins with the selection of events with exactly one charged lepton, either an electron with transverse momentum
$p_{T}>15$~GeV and pseudorapidity~\cite{defseta} $|\eta|<$ 1.1 or $1.5 < |\eta| < 2.5$ or a muon with $p_{T}>15$~GeV and $|\eta| < 2.0$. 
Events are also required to have \MET~$>15$~(20)~GeV for the electron (muon) channel and two or three jets with $p_{T}>20$~GeV 
(after calibration of the jet energy~\cite{Abazov:2011vi}) and $|\eta|<2.5$. 
\MET\ is calculated from the energy deposits in the
calorimeter cells and is corrected for the presence of muons~\cite{Abazov:2012wh}. 

Electron candidates are identified based on a multivariate discriminant that
uses information from the central tracker, preshower detectors, and calorimeter.
Muon candidates are identified from the hits in the muon system that 
are matched to a central track
and must be isolated from the energy deposits in the calorimeter.
Inefficiencies introduced by lepton identification and isolation criteria
are determined from $Z\to\ell\ell$ data and used to correct the efficiency in simulated events to match that in the data.

Jets are reconstructed by using a midpoint cone algorithm~\cite{Blazey:2000qt} with a radius
of $\Delta\mathcal{R} =\sqrt{(\Delta y)^2 + (\Delta \phi)^2} =0.5$, where $y$ is the rapidity.
Differences in efficiency for jet identification and jet energy
resolution between the
data and simulation are applied as corrections to the MC~\cite{Abazov:2012wh}.

Comparison of {\sc alpgen} with other generators~\cite{Alwall:2007fs} and with the data~\cite{Abazov:2008ez}
shows discrepancies in distributions of lepton and jet $\eta$,
dijet angular separations, and the $p_T$ of $W$ and $Z$ bosons for $V$+jets events.
The data are therefore used to correct the {\sc alpgen} $V$+jets MC events by weighting 
the simulated distributions of lepton $\eta$, leading and second-leading jet $\eta$, 
$\Delta\mathcal{R}$ between the two leading jets, and the $W$ boson $p_T$ 
through the use of functions that bring the total simulated background into agreement with the data before $b$ tagging,
similar to the method employed in Ref.~\cite{Abazov:2012wh}.

Multijet backgrounds are estimated from the data~\cite{Abazov:2012wh}.
Before applying $b$-tagging, we perform a
fit to the distribution of the transverse mass~\cite{Abazov:2012bv} of the 
$W$ boson candidate ($M_T^W$)
to determine the normalization of the MJ and $V$+jets backgrounds simultaneously.
To suppress MJ background, events with \mbox{$M_T^W < (40 - 0.5\times\MET)$ }
are removed in both the electron and muon channels.

To further suppress the MJ background, we construct a multivariate discriminator
that exploits kinematic differences between the MJ background and signal.  
The multivariate disciminator is a boosted decision tree (BDT) implemented in the {\sc tmva} package~\cite{Hocker:2007ht}.  
The output distribution in the data is well modeled by
the total expected simulated and MJ backgrounds
and is used as one of the inputs to the final signal discriminant.

The $b$-tagging algorithm for identifying jets originating from $b$ quarks is 
based on a combination of variables sensitive to the presence of tracks
or secondary vertices displaced significantly from the $p\bar{p}$ interaction vertex. 
This algorithm provides improved performance over
an earlier neural network algorithm~\cite{Abazov:2010ab}.
The efficiency is determined for taggable jets, which contain at least two tracks with each having
at least one hit in the silicon microstrip tracker. 
The efficiency for jets to satisfy the taggability and 
$b$-tagging requirements 
in the simulation is corrected to reproduce the data.

Events must have at least one $b$-tagged jet.
If exactly one jet is $b$-tagged, the $b$-identification discriminant output of that jet must satisfy 
the tight selection threshold described below.
Such events are classified as having one tight $b$ tag.
Events with two or more $b$-tagged jets are assigned to either the two loose $b$ tags, two medium $b$ tags, or two tight $b$ tags category,
depending on the value of the average $b$-identification discriminant of the two jets with the highest discriminant values.
The operating point for the loose (medium, tight) threshold has an
identification efficiency of 79\%\ (57\%, 47\%) for individual $b$ jets, averaged over selected jet $p_T$ and $\eta$ distributions, 
with a $b$-tagging misidentification rate of 11\%\ (0.6\%, 0.15\%) for light-quark jets, calculated by the method described in Ref.~\cite{Abazov:2010ab}. 

\begin{figure*}[htbp]
{
\centering {
\includegraphics[width=0.32\textwidth]{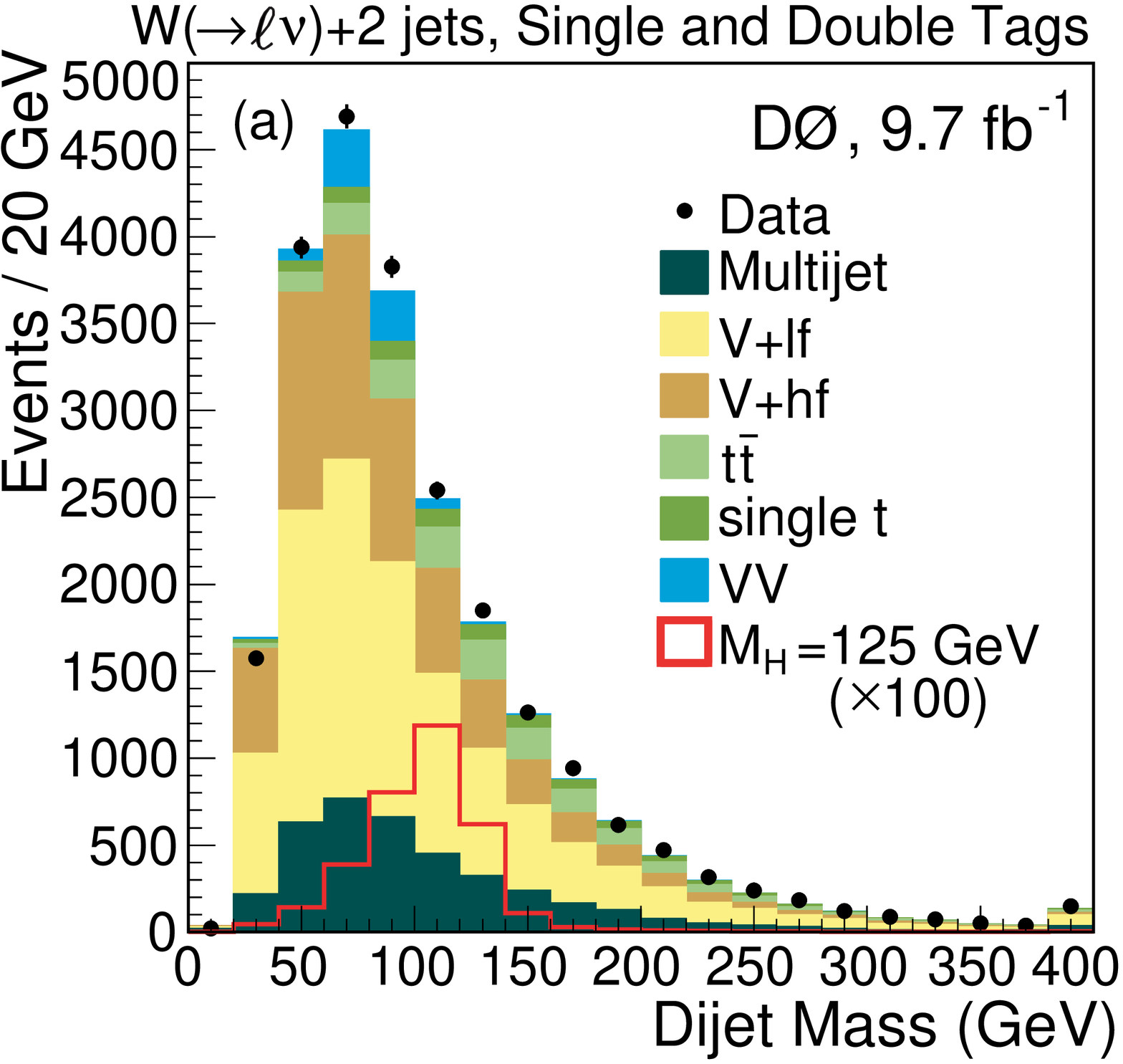}
\includegraphics[width=0.32\textwidth]{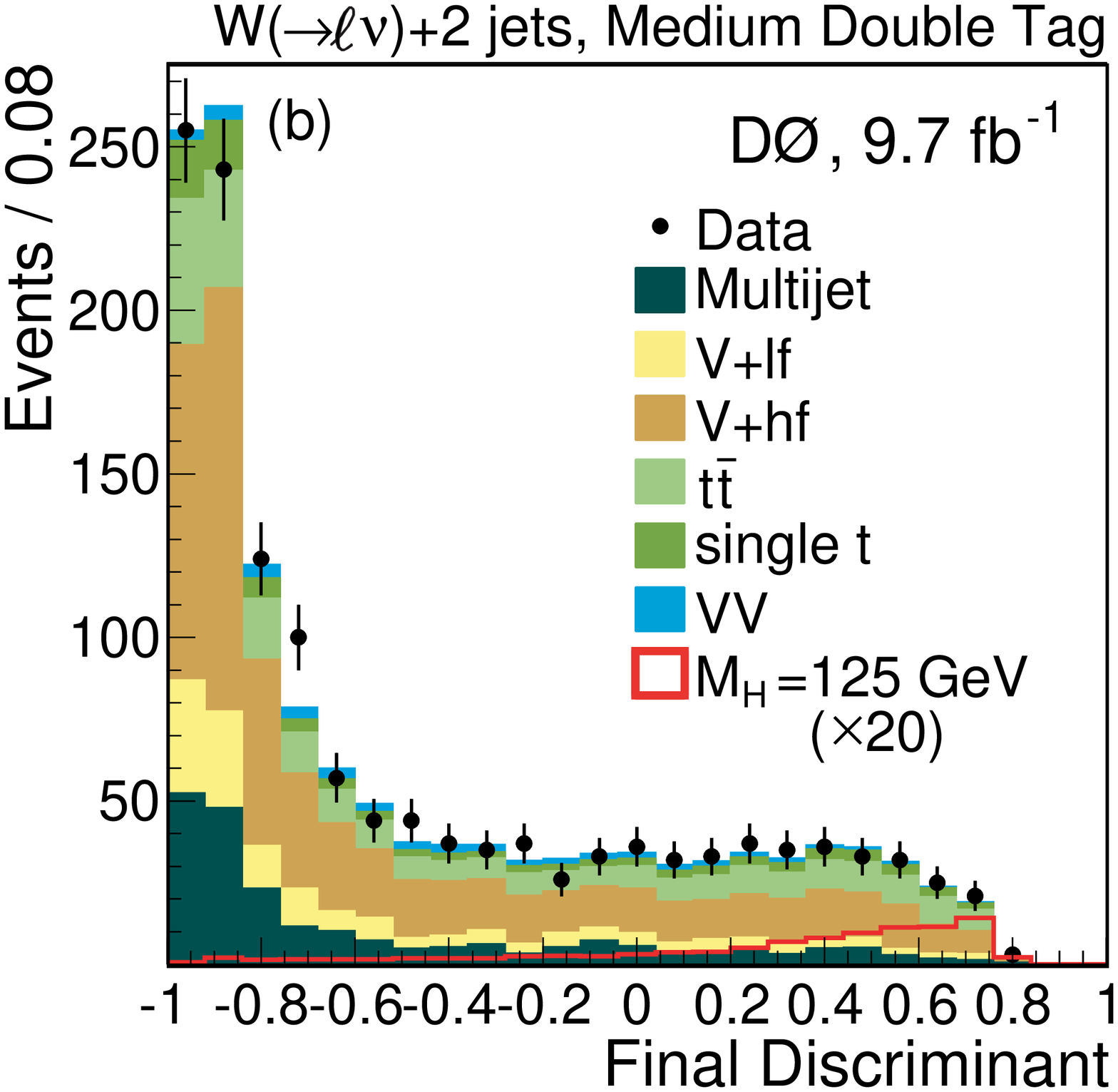}
\includegraphics[width=0.32\textwidth]{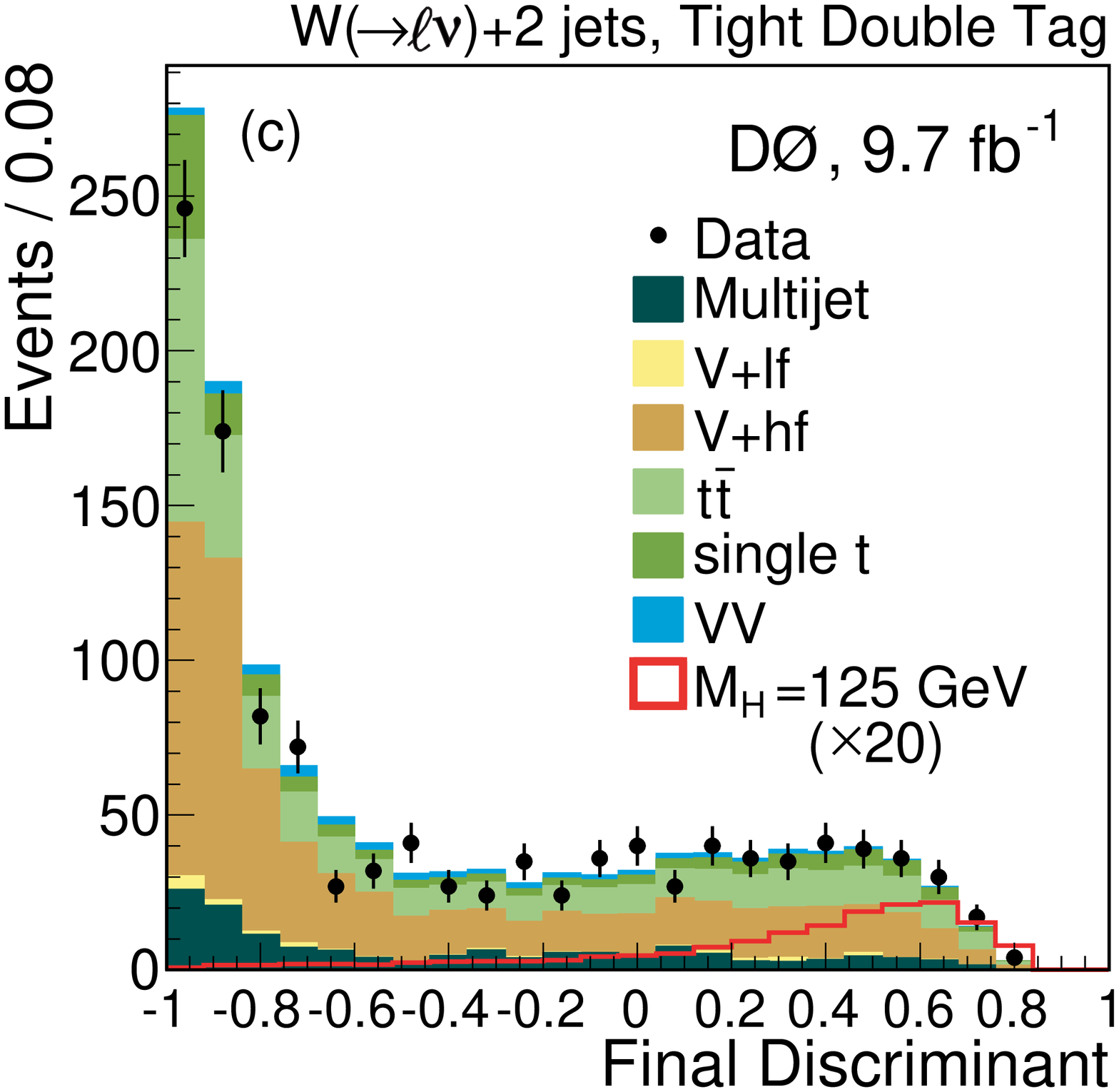} 
} \caption{\label{fig:2jMVA}
(color online). (a) The dijet mass distribution for all $b$-tag categories and two-jet exclusive events. 
(b) The final BDT output for two medium $b$-tagged events and (c) two tight $b$-tagged events.
Electron and muon channels are combined.
The Higgs boson signal is shown for $M_H=125$~GeV.
Signal events are scaled by a factor of 100 in (a) and 20 in (b) and (c).
} }
\end{figure*}

After applying these selection criteria, 
the expected event yields for the backgrounds and for a Higgs boson with mass $M_{H}=125$~GeV 
are compared to the observed number of events in Table~\ref{tab:yields}.
Figure~\ref{fig:2jMVA}(a) shows the distribution of the dijet invariant mass,
using the two jets with the highest $b$-identification output, for events with exactly two jets and all $b$-tagged categories.
The data are well described by the predicted background in all $b$-tag categories.

\begin{table}[htb]
\begin{center}
\caption{\label{tab:yields} {
Summary of event yields for $W$+2 and $W$+3 jets final states.
The number of events in the data is compared with the expected number of background  events.
Signal contributions ($M_H=125$~GeV) are shown for $WH$ and $ZH$ production with $H\to b\bar{b}$.
All listed signal sources are considered when setting limits.
Uncertainties include both statistical and systematic contributions, as described later in this Letter.
}}
\begin{tabular}{lr@{$\,\pm\,$}lr@{$\,\pm\,$}lr@{$\,\pm\,$}l}
\hline
\hline
 & \multicolumn{2}{c} { Pre-$b$-tag } & \multicolumn{2}{c} { One tight $b$-tag  } & \multicolumn{2}{c} { Two $b$-tags } \\
\hline
$WH$            &	41.2	&	3.2	&	12.5	&	1.2	&	17.3	&	1.7		\\
$ZH$           	&	4.7	&	0.4	&	1.4	&	0.1	&	1.9	&	0.1		\\
\hline
$VV$            &	6824	&	678	&	648	&	55	&	256	&	18		\\
$V$+lf          &	206 358  &	18 624	&	7149	&	794	&	2527	&	306		\\
$V$+hf	        &	34 068	&	4447	&	6486	&	1510	&	3164	&	739		\\
Top	        &	7222	&	555	&	2413	&	229	&	2437	&	238		\\
Multijet	&	68 366	&	6668	&	4634	&	473	&	2020	&	192		\\
\hline
All bkg.        &	322 838  &	24 756	&	21 330	&	2190	&	10 404	&	1059		\\
Data	        & \multicolumn{2}{c}{	322 836} & \multicolumn{2}{c}{	20 684}  & \multicolumn{2}{c}{	10 071	}	\\
\hline
\hline
\end{tabular}
\end{center}
\end{table}

To separate the signal and background, we use final BDTs trained on the $WH\to\ell\nu b\bar{b}$ signal samples and all the SM processes as background.
We train an independent final BDT, using an individually optimized set of inputs, for each lepton flavor, jet multiplicity, $b$-tag category, and $M_H$ value considered, with $M_{H}$ varying between 100 and 150~GeV in 5~GeV steps.
When selecting input variables, we ensure that each is well modeled and displays good separation between the signal
and one or more backgrounds.  
Figures ~\ref{fig:2jMVA}(b) and \ref{fig:2jMVA}(c) shows the final BDT output distributions for the two medium and two tight $b$-tag channels in two-jet events with electron and muon channels combined.

Uncertainties on the normalization and shape of the final BDT output distributions affect our sensitivity to a potential signal.
Theoretical uncertainties include uncertainties on 
the $t\bar{t}$ and single top quark production cross sections (each having a 7\%\ uncertainty \cite{Langenfeld:2009wd,Kidonakis:2006bu}),
an uncertainty on the diboson production cross section (6\% \cite{Campbell:1999ah}), $V$+light-flavor production (6\%), and $V$+heavy-flavor production (20\%, estimated from {\sc mcfm}~\cite{Campbell:2001ik,mcfm_code}).

Uncertainties from modeling that affect both the shape and normalization of the final BDT distributions include uncertainties 
on trigger efficiency as derived from the data (3\%--5\%),
lepton identification and reconstruction efficiency (5\%--6\%),
reweighting of {\sc alpgen} MC samples (2\%), and the MLM matching~\cite{Mangano:2002ea} applied to $V$+light-flavor events ($\approx0.5$\%).
Uncertainties on the {\sc alpgen} renormalization and factorization scales are evaluated by multiplying the nominal scale 
for each, simultaneously, by factors of 0.5 and 2.0 (2\%),
while uncertainties on the choice of parton distribution functions (2\%) are estimated by using the prescriptions of Ref.~\cite{Pumplin:2002vw,Stump:2003yu}.

Experimental uncertainty that affects only the normalization of the expected signal and simulated backgrounds arises from the uncertainty on 
the integrated luminosity (6.1\%)~\cite{Andeen:2007zc}.  Those that affect the 
final BDT distribution shapes include jet taggability (3\% per jet), 
$b$-tagging efficiency (2.5\%--3\%\ per heavy-quark jet),
the light-quark jet misidentification rate (10\% per jet), jet identification efficiency (5\%), 
and jet energy calibration and resolution (varying between 5\%\ and 15\%, depending on the process 
and channel).   The MJ background model has a contribution from the statistical uncertainty of 
the data after tagging (10\%--20\%). 

To 
demonstrate measurement of processes with small
cross sections in the same final state as $WH$, we train a discriminant with $WZ$ and $ZZ$ production as the signal,
using the same event selection and input variables.  We observe a 
\dibosig\ s.d.~excess in the data over the background expectation,
and our expected sensitivity is \diboexpsig\ s.d. 
If interpreted as a cross section measurement, the resulting scale factor 
with respect to the predicted SM value~\cite{Campbell:1999ah,mcfm_code} 
of \diboSMxsec\ is \diboxsec.

\begin{figure}[htb]
{
\centering {
\includegraphics[width=0.43\textwidth]{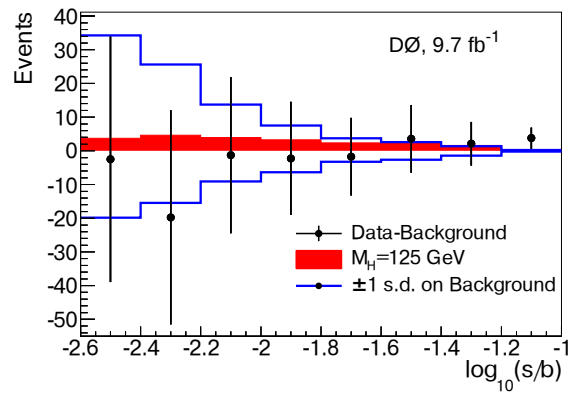}
} \caption{\label{fig:bsd} 
(color online). Distribution of the difference between the data and background expectation
of the final BDT discriminant output for $M_{H}=125$~GeV for the background-only model, shown with statistical uncertainties (points with error bars). 
The solid lines represent the $\pm$1 s.d.\ systematic uncertainty after constraining with the data.
The darker shaded region is the expected final BDT distribution for a SM Higgs signal for $M_H=125$~GeV.  Here we combine BDT discriminant bins
from each channel according to the bins' $\log_{10}(s/b)$ values. 
} }
\end{figure}

\begin{figure}[htb]
{
\centering {
\includegraphics[width=0.43\textwidth]{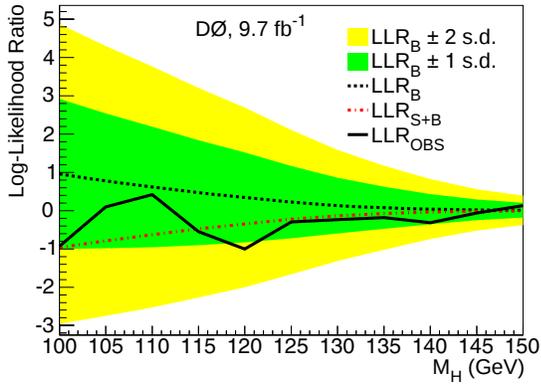} 
} \caption{\label{fig:limits} 
(color online). Log-likelihood ratio for the background-only model (LLR$_B$, with 1 and 2~s.d.\ uncertainty bands), 
signal+background model (LLR$_{S+B}$) and data (LLR$_{\text{obs}}$) versus $M_{H}$.
} }
\end{figure}

In the search for the SM Higgs boson we observe no significant excess relative to the SM expectation and proceed to set upper limits on the SM Higgs boson production cross section.
We calculate all limits at the 95\%\ C.L.\ using the modified frequentist CL$_s$ approach with a Poisson log-likelihood ratio
of the signal+background hypothesis to the background-only hypotheses
as the test statistic~\cite{Junk:1999kv,Read:2002hq,wade_tm}.  We treat systematic uncertainties 
as ``nuisance parameters'' constrained by their priors, and the best fits of these
parameters are determined at each value of $M_{H}$ by maximizing the
likelihood with respect to the data. We remove the $V$+jets normalization obtained from the  $M_T^W$ distribution and allow the components to vary
by the aforementioned uncertainties of 6\%\ and 20\%\ on $V$+light-flavor and $V$+heavy-flavor production, respectively.  
Independent fits are performed to the background-only and signal-plus-background
hypotheses. All correlations are maintained among channels and between the signal and background.  Figure \ref{fig:bsd} shows the 
background-subtracted data along with the best fit 
for the background-only model
$\pm 1$~s.d.\ systematic uncertainties and the expected signal contribution 
for all channels combined,
where we combine bins from each channel according to their $\log_{10}(s/b)$ value in order to group bins with similar sensitivity.
The log-likelihood ratios for the background-only model and the signal-plus-background model 
as a function of $M_{H}$ are shown in Fig.~\ref{fig:limits}.
The upper limit on the cross section 
for $\sigma( p\bar{p} \rightarrow H+X) \times \mathcal{B}(H \rightarrow b \bar{b})$
for $M_{H}=125$~GeV is a factor of \obslimA\ larger than the SM expectation
and our expected sensitivity is \explimA. 
The corresponding observed and expected limits relative to the SM expectation
are given in Table~\ref{tab:limits}.

\begin{table}[htb]
{
\begin{center}
\caption{\label{tab:limits}
The ratio of the observed, $R_{obs}$, and expected, $R_{expt}$, 
95\% upper limit to the SM Higgs boson production cross section.}
{ \renewcommand{\arraystretch}{0.8}
\begin{tabular}{lcccccccccccc}
\hline
\hline
$M_H$ (GeV)& 100  &  105 &  110 &  115 &  120  &  125 &  130 &  135 &  140  &  145  &  150  \\
\hline
$R_{expt}$ & 2.2 & 2.5 & 2.9 & 3.2 & 3.8 & 4.7 & 6.2 & 8.2 & 11.7 & 17.5 & 25.6 \\
$R_{obs} $ & 2.8 & 2.6 & 2.9 & 3.7 & 5.0 & 5.2 & 6.8 & 8.9 & 15.1 & 18.8 & 21.8 \\
\hline
\hline 
\end{tabular}
}
\end{center}
}
\end{table}

In conclusion, we have performed a search for SM Higgs boson production in $\ell$+\MET+jets final states 
using two or three jets and $b$-tagging
with the full run II data set of 9.7 fb$^{-1}$ of integrated luminosity from the D0 detector.
The results are in agreement with the expected event yield, 
and we set upper limits 
on $\sigma(p\bar{p} \rightarrow H+X) \times \mathcal{B}(H \rightarrow b \bar{b})$
relative to the SM Higgs boson cross section $\sigma_{\rm SM}$ for $M_H$ between 100 and 150~GeV,
as summarized in Table~\ref{tab:limits}.
For $M_{H}=125$~GeV, the observed limit normalized to the SM prediction is \obslimA\ and the expected limit is \explimA.  

%
We thank the staffs at Fermilab and collaborating institutions,
and acknowledge support from the
DOE and NSF (USA);
CEA and CNRS/IN2P3 (France);
MON, NRC KI and RFBR (Russia);
CNPq, FAPERJ, FAPESP and FUNDUNESP (Brazil);
DAE and DST (India);
Colciencias (Colombia);
CONACyT (Mexico);
NRF (Korea);
FOM (The Netherlands);
STFC and the Royal Society (United Kingdom);
MSMT and GACR (Czech Republic);
BMBF and DFG (Germany);
SFI (Ireland);
The Swedish Research Council (Sweden);
and
CAS and CNSF (China).

\bibliography{higgs}

\begin{thebibliography}{10}
\bibitem{defseta}
The pseudorapidity $\eta = - \ln \left[ \tan\frac{\theta}{2}\right]$, where
  $\theta$ is the polar angle as measured from the proton beam axis.
\end{thebibliography}
\bibliographystyle{h-physrev3}

\end{document}